# Polarization-Encoded BB84 QKD Transmitter Sourced by a SiGe Light Emitter

**Florian Honz[(1)], Nemanja Vokić[(1)], Philip Walther[(2)], Hannes Hübel[(1)] and Bernhard Schrenk[(1)]**

[(1)]AIT Austrian Institute of Technology, Center for Digital Safety&Security / Security & Communication Technologies, 1210 Vienna, Austria.

[(2)]University of Vienna, Faculty of Physics, 1090 Vienna, Austria

Author e-mail address: florian.honz@ait.ac.at

**Abstract:** We demonstrate a polarization-encoded BB84 transmitter sourced by a SiGe light source and show that such a potentially "all-silicon" QKD scheme can operate well below the QBER threshold at which secret keys can be established. © 2023 The Author(s)

## 1. Introduction

Quantum Key Distribution (QKD), a technology that guarantees the absolute secure exchange of symmetric cryptographic keys, has reached full technological maturity. 19-inch compatible QKD devices have been readily installed and operated in the last years in many deployed network configurations around the world. However, for a ubiquitous roll-out in scenarios where a very large number of devices are needed, QKD systems should feature a simplified hardware implementation and have to be produced much smaller in size and at reduced costs, as investigated in recent photonic-integrated QKD realizations [1-3]. A representative deployment scenario for such simplified QKD systems is the intra-datacenter environment. The high server density of more than 10,000 per datacenter [4], paired with the advent of zero-trust models associated to security within these information hotspots, make datacenter networks a new battleground for QKD. However, the requirements for this environment are more stringent than for any other traditional network domain: Simplicity for the involved sub-systems prevails and must run down-rack. Fortunately, the continuous advances in opto-electronic integration technologies have enabled highly compact transceivers, which can unleash new paradigms such as co-packaged optics, where opto-electronics and electrical ASICs are tightly shoehorned in the same package. To introduce QKD in this challenging domain by means of further functional expansion, a similar degree of integration density and simplicity needs to be accomplished for the QKD sub-systems. Although these efforts can build on silicon integration, not all constituent elements can be realized in a monolithic fashion – especially not the light source.

In this work, we employ a SiGe light emitter to source a BB84 polarization encoder as a first feasibility study towards an all-silicon QKD transmitter. We demonstrate, for the first time to our best knowledge, that such a QKD scheme operates at a QBER of 7.7% and thus below the threshold at which a secret key can be established.

## 2. Towards an All-Silicon QKD Transmitter for Deployment with High Asymmetrical Complexity

Advantageously, the quantum channel associated to QKD is unidirectional. With this, QKD can employ an asymmetrical complexity profile in transmitter and receiver sub-systems, in a way that complex elements (such as Bob's SPADs) are centralized while simpler elements (such as Alice' transmitter) remain distributed in a network. This case is highlighted in Figure 1 in view of a datacenter pod. QKD networks might span within a rack, where a top-of-the-rack (ToR) Bob is co-located with the ToR switch and exchanges a key with many rack-side Alices, taking advantage of optical lightpath switching – a technology that is expected to soon inundate datacenter networks [5]. Similarly, Bob can be centralized for the entire pod, so that complexity is further pooled at a single point of the network. Therefore, Alice' QKD transmitter becomes the prime focus of complexity reduction, for which this work lays the groundwork for an all-silicon BB84 encoder by introducing a SiGe light emitter that is compatible with silicon-on-insulator platforms. It feeds a polarization-encoding BB84 transmitter that encodes the light in two bases (horizontal/vertical, H/V, and diagonal/anti-diagonal, D/A), whose signal is then launched at the quantum level to a centralized Bob. Such a polarization-encoded

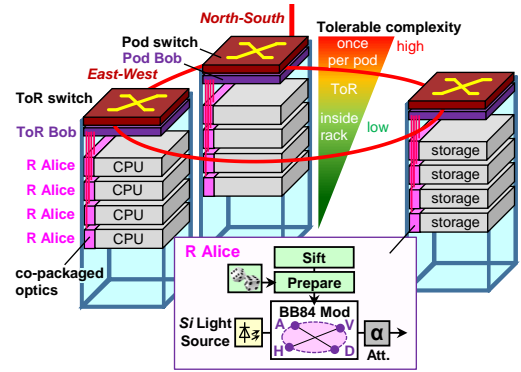

Fig. 1. Datacenter pod with low-complexity distributed Alice QKD transmitters aiming at an all-silicon layout.

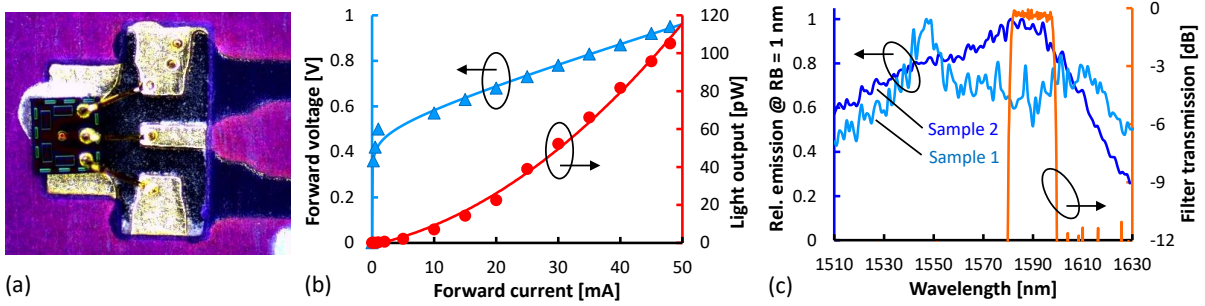
Fig. 2. (a) SiGe emitter used as Alice' light source and its (b) VLI characteristics and (c) emission spectrum for two emitter samples.

protocol can be readily implemented on silicon, as demonstrated in [6, 7]. Here, we complete the functional toolbox of an all-silicon based Alice by demonstrating BB84 in combination with a silicon emitter.

## 3. Silicon Light Emitter

Figure 2a shows the die-level light emitter that we employed in this experiment. It is a forward-biased, vertically fiber-coupled SiGe PIN junction whose VLI characteristics are reported in Fig. 2b. The *V-I* characteristics (▲) resemble that of a diode, with a forward voltage of less than 1 V. The corresponding *L-I* performance (●) shows a threshold-less behavior that reaches a power of 100 pW at a current of 46 mA. This level corresponds to an optical power of -70 dBm, which, for a typical symbol rate of 1 GHz of a QKD transmitter, is above the single photon-level. This leaves a headroom of 10 dB to account for optical losses at the transmitter. The optical emission spectrum is presented in Fig. 2c for two different emitter dies. We notice a LED-like behavior for the SiGe emitters, as it can be expected for light sources based on simple junctions. For the second sample, which we later used in the experiment, the peak emission wavelength is 1581 nm. For both samples, the emission is found within the C+L band range and thus centric to the component spectrum of telecom technology.

## 4. Experimental Setup

The experimental setup is shown in Fig. 3a. We have chosen a BB84 encoder that emits the four target states by switching the H/V/D/A polarizations. Four Mach-Zehnder modulators (MZM) encode states in the H/V and D/A bases, defined by subsequent polarization beam combiners (PBC) and free-space optics that set the rotation of the D/A basis with respect to H/V through a half-wave plate (HWP). Pulse carving is integrated with the MZM drive to suppress symbol transitions. An attenuator sets the signal launch at the quantum level with a mean photon number of $\mu_Q = 0.1$ photons/pulse. The implementation of a full decoy state protocol was not attempted in this work.

For the light emitter that sources the MZMs, we employed the proposed SiGe emitter (σ) and the amplified spontaneous emission (ASE, α) of an Erbium-doped fiber amplifier (EDFA). An optical bandpass filter (BPF) was optionally added to evaluate the QKD performance for different bandwidth settings of the light source and a polarization beam splitter (PBS) was used to define a clean polarization state at the input of the MZMs.

At Bob, we use manual polarization control (PC) to optimize the input state of the BB84 analysis module, which measures the received states in the H/V and D/A bases. Due to a limited inventory of SPADs, we used a pair of free-running InGaAs SPADs with dark count rates of 560 and 525 cts/s to consecutively measure in the H/V and D/A bases. A time-tagging module (TTM) then registers the detection events for a real-time QBER evaluation, which includes frame synchronization with a pre-defined pseudo-random bit sequence applied at Alice' transmitter, temporal filtering within 50% of the symbol period to cut part of the dark count noise and basis sifting.

The measurements at the quantum level were accompanied by characterization and functional verification at the classical level, for which the modified

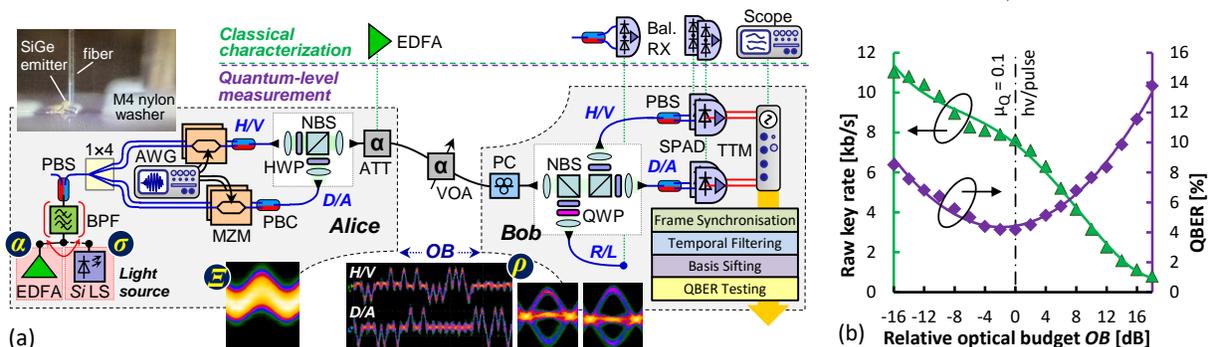
Fig. 3. (a) Experimental setup and (b) raw key rate and QBER as function of the optical budget for the ASE-sourced BB84 transmitter.

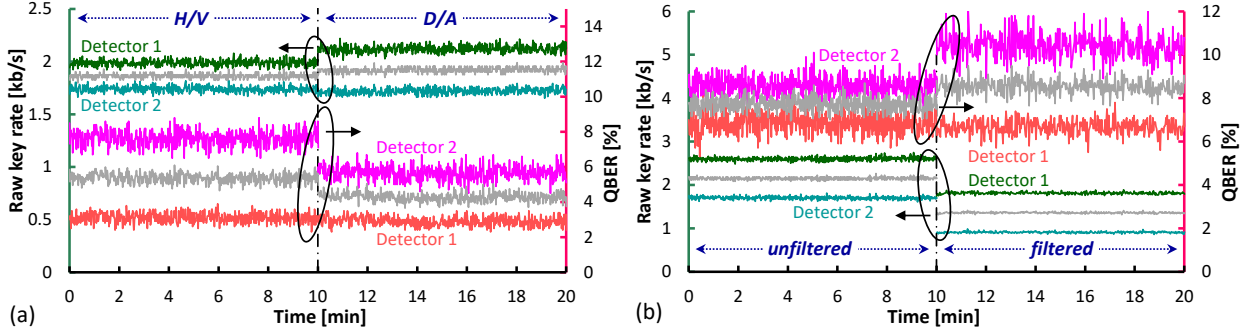

Fig. 4. Short-term evolution of the raw key rate and QBER for (a) ASE-sourced and (b) SiGe emitter sourced BB84 transmitter.

elements in the setup are highlighted in Fig. 3a, together with the carved optical output at Alice ($\Xi$) and the characteristic 3-level signals ($\rho$) received at Bob for the 4-state BB84 encoding.

## 5. Results and Discussion: Raw Key Rate and QBER

As a reference, we first evaluated the BB84 encoder driven by ASE ($\alpha$ in Fig. 3a) sliced by a BPF with a bandwidth of 25 GHz at 1539.1 nm. To obtain the sensitivity of the system to optical loss, we added a variable optical attenuator (VOA) between Alice and Bob to set the optical budget (OB) for transmission. Figure 3b presents the raw key rate (▲) and the QBER (♦) as function of the OB, for which positive (extra loss) but also negative ($\mu > \mu_Q$) values have been set – the latter for characterizing the saturation threshold of the SPAD receivers.

At OB = 0 dB, for which $\mu = \mu_Q$, we obtain a raw key rate of 7.6 kb/s at a QBER of 4.2%. The QBER threshold of 11%, which we consider the cut-off for generating a secret key [8], is reached after accommodating an OB of 15.2 dB. This would practically allow to accommodate loss due to an optical switch that acts as signal distributing element between one Bob and $N$ Alices in a datacenter pod, or even a passive 1:$N$ split for $N$ up to 16.

Figure 4a presents the evolution of the QBER and raw key rate for the two constituent detectors of the H/V and D/A bases that have been measured consecutively over a short time of ~10 min and at $\mu = \mu_Q$. The average QBER for the H/V and D/A basis was 5.37% (3$\sigma$ = 0.78%) and 4.28% (3$\sigma$ = 0.70%), respectively, while the raw key rates are 1.87 and 1.92 kb/s. This proves the correct operation of the BB84 transmitter when sourced by incoherent light.

We then switched to the SiGe emitter ($\sigma$ in Fig. 3a) as the optical feed for Alice' BB84 encoder. Figure 4b again presents the time evolution for $\mu = \mu_Q$ when the emission of the SiGe light source is unfiltered and filtered by a 14-nm wide bandpass filter, whose transmission is reported in Fig. 2c. For the same temporal filtering ratio of 50%, the average QBER without and with filtering is 7.71% and 8.53% at raw key rates of 2.15 and 1.36 kb/s, respectively. We do not see an improvement due to a reduction in spectral width of the seed light for this back-to-back investigation, but expect a reduction in transmission-related impairments such as chromatic dispersion or dispersion-induced increase in relative intensity noise when transmitting over intra-datacenter fiber reach. The accomplished QBER clearly demonstrates the feasibility of realizing an all-silicon QKD transmitter in near future.

## 6. Conclusion

We have experimentally evaluated the use of a silicon-based light source to feed a QKD transmitter. The employed SiGe emitter features a C+L band light emission at the single-photon level for QKD transmitters pulsed at 1 GHz. Evaluation for polarization-encoded BB84 showed an acceptable performance with a QBER of <5.4% at a raw key rate of >3.7 kb/s for an ASE-sourced QKD transmitter, while the SiGe emitter sourced transmitter performed at 7.7% and thus well below the QBER threshold at which a secret key can be established. The results prove the feasibility of employing silicon light sources for QKD applications, where stringent cost requirements apply and where the small footprint and low manufacturing complexity of all-silicon QKD transmitters are highly beneficial.

Acknowledgement: This work was supported by the ERC under the EU Horizon-2020 programme (grant agreement No 804769) and by the Austrian FFG Research Promotion Agency and NextGeneration EU (grant agreement No FO999896209).